

\magnification=1200
\hsize = 5.5 true in
\hoffset = 0.5 true in
\baselineskip=20pt
\lineskiplimit=1pt
\lineskip=2pt plus .5pt
\parskip = 3pt

\def\today{\number\month /\number\day /\number\year}
\pageno=0
\headline={\rm \ifnum \pageno<1 LA-UR-93-4019 \hfil submitted to
	{\it Molec. Phys.\/} \today
		\else LA-UR-93-4019 \hfil
		``Electron Density Functional Models''
			\hfil \today \fi}
\footline={\rm \ifnum \pageno<1 \hfil \else \hfil \folio \hfil \fi}

\topskip = 48pt
\parindent =  20pt
\vfil
\centerline{\bf Comparison of Electron Density Functional Models}

\bigskip
\bigskip

\centerline{
		\it Gary G. Hoffman
		\footnote\dag{	Department of Chemistry,
				Florida International University,
				Miami, FL 33199
				}
		and Lawrence R. Pratt
		\footnote*{	Los Alamos National Laboratory,
				Los Alamos, NM 87545.
				}
		}
\bigskip
\bigskip


\parindent =  20pt
\topskip = 10pt

\beginsection{Abstract}

Presented here are calculations of the distortion of the density of
an electron gas due to the electrostatic field of a proton.  Several
models based upon the local density approximation (LDA) of density
functional theory [linear response theory, Kohn-Sham (KS), optimized
Thomas-Fermi theory (OTF), and OTF plus perturbation corrections] are
compared with one another.  These models, in turn, are compared with
available results of quantum Monte Carlo (QMC) calculations for the
same system.  Comparison of the KS results with the OTF results shows
a very reasonable agreement that seems to be progressively improvable.
This provides encouragement for the application of the OTF model to
condensed phase systems.  The QMC calculations of the density do not
agree well with the density functional results.  The reasons for this
poor agreement are not clear.  This particular system is expected to
be an ideal one for application of the LDA and, thus, the poor
agreement is of fundamental importance.  The lack of detail presented
in the available QMC results leads us to conclude that the QMC
calculations should be attempted again.

\vfil \eject

published as:  {\it Molec. Phys.\/} {\bf 82}, 245-261(1994)

\end